\documentclass{ws-rv10x7}
\usepackage{ws-rv-van}     
\usepackage{ws-rv-thm}     
\usepackage{subfigure}
\usepackage{xcolor}
\makeindex
\begin{document}

\chapter[The Kauzmann Transition to an Ideal Glass Phase]{The Kauzmann Transition to an Ideal Glass Phase\label{ch1}}

\author[C. Cammarota, M. Ozawa, and G. Tarjus]{Chiara Cammarota$^*$, Misaki Ozawa$^\dagger$, and Gilles Tarjus$^\ddagger$}

\address{$^*$Dipartimento di Fisica, La Sapienza Universit\'a di Roma, P.le A. Moro 5, 00185 Rome, Italy}

\address{$^\dagger$Laboratoire de Physique de l’Ecole Normale Sup\'erieure, ENS, Universit\'e PSL, CNRS, Sorbonne Universit\'e, Universit\'e Paris Cit\'e, F-75005 Paris, France}

\address{$^\ddagger$LPTMC, CNRS-UMR 7600, Sorbonne Universit\'e, 4 Pl. Jussieu, F-75005 Paris, France}

\begin{abstract}
The idea that a thermodynamic glass transition of some sort underlies the observed glass formation has been highly debated since Kauzmann first stressed 
the hypothetical entropy crisis that could take place if one were able to equilibrate supercooled liquids below the experimental glass transition temperature 
$T_g$. This {\it a priori} unreachable transition at some $T_K<T_g$ has since received a firm theoretical basis as a key feature predicted by the 
mean-field theory of the glass transition. In this chapter, we assess whether,  and in which form, such a transition can survive in finite dimensions, and we 
review some of the recent computer simulation work addressing the issue in $2$- and $3$-dimensional glass-forming liquid models. We also discuss 
theoretical reasons to focus on an apparently inaccessible singularity.
\end{abstract}

\date{\today}



\body

\section{Introduction}

The existence of a thermodynamic phase transition underlying the experimentally observed glass transformation phenomenon has been 
a recurring theme in theoretical studies of glass formation. This transition would be between the 
liquid --most generally a supercooled liquid phase which is metastable with respect to some stabler crystalline phase--  and an ideal glass phase. The issue may sound just as futile as arguing over how many angels can dance on the 
head of a pin because the thermodynamic transition is unreachable due to the very strong slowing down of relaxation that is precisely the 
phenomenological fingerprint of glass formation. Yet, one can argue that there is some merit to trying to address the problem. One first obvious 
reason is that despite being unobservable, the putative thermodynamic transition may still control the physics of the glassy slowdown and provide a 
framework and scaling laws to describe the empirical data. Accordingly, more or less indirect signatures or vestiges of the thermodynamic transition 
could be probed and would allow for distinguishing  the theoretical approaches that do and do not predict such features. This chapter is devoted to a discussion of this issue.
\\

The story of an unreachable transition at some temperature $T_K$ lower than the experimental glass transition (or transformation) temperature $T_g$ 
begins with W. Kauzmann's seminal paper in 1948~\cite{kauzmann1948nature}. Addressing the nature of metastability and of the glass transition, 
Kauzmann collected equilibrium entropy data from various glass-forming materials and plotted the temperature dependence of the entropy difference 
between the supercooled liquid and the crystal. He noted that this difference decreases sharply with decreasing temperature, more so for molecular 
glass-forming liquids now known as "fragile"~\cite{angell1995formation} such as glucose or lactic acid for which the entropy difference drops by a factor of $2$ or 
$3$ between the melting point and $T_g$. At $T_g$ the entropy difference essentially saturates because the liquid falls in a nonequilibrium glass state. 
However, one may wonder what would happen if one were able to equilibrate the liquid to still lower temperatures, a point that Kauzmann 
qualitatively illustrated by extrapolating the experimental curve below $T_g$. He found that a simple extrapolation leads to an apparent paradox: The entropy difference between the liquid and the crystal vanishes at a nonzero temperature $T_K<T_g$ below which it becomes negative. This temperature 
$T_K$ is now known as the Kauzmann temperature. 

Although Kauzmann himself preferred an interpretation in terms of a pseudo-critical point at (or above) $T_K$ marking the limit of stability of the supercooled liquid with respect to crystal nucleation, $T_K$ has since been associated in many glass studies with a thermodynamic (equilibrium) transition to an ideal glass 
phase. This view has been pursued by J. H. Gibbs and his coworkers. Gibbs and DiMarzio found through a mean-field quasi-lattice approach an  
equilibrium glass transition at which the "configurational entropy" of glass-forming polymers vanishes~\cite{gibbs1956nature,gibbs1958nature}. Adam and Gibbs later proposed a mechanism to relate the slowdown of relaxation as one lowers the temperature to the dearth of available configurations as quantified 
by the decrease of the configurational entropy. This mechanism involves cooperatively rearranging regions whose size diverges at the equilibrium glass 
transition temperature $T_K$~\cite{adam1965temperature}.
\\

What gave firmer ground to the notion of a thermodynamic glass transition at some nonzero $T_K<T_g$ is the insight by T. R. Kirkpatrick, D. Thirumalai, and P. G. Wolynes that such an entropy-vanishing transition is present in some mean-field spin-glass models, such as Potts glasses and $p$-spin models~\cite{kirkpatrick1987connections,kirkpatrick1987p}. The transition, which in the replica formalism corresponds to a 
1-step replica symmetry breaking (RSB)~\cite{mezard1987spin}, has been dubbed "random first-order transition", and its scaling extension to finite dimensions has been developed in a series of papers starting with Ref.~\cite{kirkpatrick1989scaling} (see also Chapter by P. G. Wolynes). In 
what follows we will often simply refer to the thermodynamic ideal glass transition as the "Kauzmann transition" to stress the fact that it corresponds to a (configurational) entropy crisis, despite the caveat concerning Kauzmann's own interpretation. 

The key step reinforcing the theoretical foundation was the full solution of glass formation for liquids in infinite dimensions that was proven to be associated with the very same 1-step replica symmetry breaking (RSB) phenomenology as the above-cited mean-field spin-glass models~\cite{charbonneau2017glass,parisi2020theory}. The Kauzmann (random first-order) transition is thus a direct prediction of the mean-field 
theory of glass-forming liquids as obtained in infinite dimensions. This recent result by itself forces one to seriously consider the issue of an underlying thermodynamic glass transition in finite dimensions. In particular, as in the conventional Landau-Ginzburg-Wilson theoretical approach of phase transitions starting from 
a mean-field description~\cite{zinn2021quantum}, it opens a line of research concerning the role of spatial fluctuations on the transition 
when moving away from infinite dimensions to reach the physical world of 2 and 3 dimensions.
\\

The goal of this chapter is to discuss the nature and the existence of a thermodynamic transition between the liquid and an ideal glass phase in finite-dimensional glass-forming liquids and, considering the fact that it is (at least currently) an essentially inaccessible transition point, to assess the usefulness of the concept for describing 
the actual glass formation process. In this respect it differs from previous literature reviewing the Kauzmann transition and arguments about its achievability~\cite{royall2018race,ediger2017perspective}.
\\

\section{Nature of the Kauzmann transition in mean-field and in finite dimensions}
\label{sec_nature}

The core of the mean-field description of glass formation is the existence of a complex free-energy landscape in which equivalent metastable states emerge in an exponentially large (in system size) number below a first critical temperature $T_d>T_K$. Below $T_d$, which corresponds to a purely dynamical transition, ergodicity is broken and the system stays trapped forever in one of the metastable states due to the presence of infinite free-energy barriers. 
As the temperature is further decreased, the number of metastable states decreases and the configurational entropy (more accurately the configurational entropy per particle), which 
is defined as the logarithm of the number of typical metastable states divided by the number of molecules, vanishes at the Kauzmann temperature $T_K$. A  
bona fide thermodynamic transition to an ideal glass then takes place. Energy, entropy, configurational entropy, and free energy are continuous at the transition but the order parameter, which similarly to spin-glass models can be best chosen as an overlap between configurations and will be discussed in more detail below, has a discontinuity between a low value (characterizing the liquid formed by the superposition of all typical metastable states) and a high value (characterizing the ideal glass). As already mentioned, this phenomenon corresponds to a 1-step RSB transition in the replica formalism (see also other chapters of this book). 

This scenario is exactly realized in liquids in infinite dimensions and can be taken as the Landau theory of the glass transition~\cite{charbonneau2017glass,parisi2020theory}. However, at odds with the treatment 
of more conventional phase transitions such as the liquid-gas one, extension of this mean-field Landau theory to include spatial fluctuations that are generically present in finite dimensions but absent in infinite dimensions is extremely delicate. One reason is that the very notion of metastability is ill-defined in finite-dimensional statistical mechanics because localized fluctuations destroy it, making the lifetime of metastable states finite in any finite dimension. Strictly speaking, the 
dynamical transition at $T_d$ cannot persist and/or must be smeared out. On the one hand, the absence of ergodicity breaking at $T_d$ is good news as the liquid is allowed to relax between $T_d$ and $T_K$ through putative activated nucleation processes~\cite{kirkpatrick1989scaling} (see also Chapter P. G. Wolynes). On the other hand, the whole theoretical construction, including the central notion of configurational entropy, may completely lose its meaning. The effect of fluctuations, which at this point we do not attempt to better characterize, may indeed be drastic and completely wipe out the mean-field 1-RSB scenario. For instance, it is known to be the case for some disordered spin 
models such as Potts glasses that show no glassiness at all or a continuous (full-RSB) transition in 
$3$ dimensions~\cite{brangian2002statics,cruz2009spin,cammarota2013fragility,takahashi2015evidence} or for the random Lorentz gas which displays a continuous localization phenomenon associated with a percolation transition in $3$ dimensions~\cite{biroli2021interplay,biroli2021mean}, both types of systems that nonetheless follow the 1-RSB scenario in the fully connected (infinite-dimensional) limit.

What about finite-dimensional glass-forming liquids? In particular, is there an operational way to define a configurational entropy? One line of research that 
started with M. Goldstein~\cite{goldstein1969viscous}, and was developed into a systematic framework by F. Stillinger and coworkers~\cite{stillinger1982hidden,stillinger1995topographic}, considers the potential energy landscape and its multiple minima, also called inherent structures. Contrary to the multiplicity of metastable states in a free-energy landscape at a nonzero temperature, the multiplicity of inherent structures is rigorously defined even in finite dimensions and their number can be used to define a configurational entropy~\cite{sciortino1999inherent}. Stillinger argued that due to the ubiquitous presence of point defects, this configurational entropy cannot vanish at a nonzero $T_K$, thereby preventing the existence of an ideal glass transition~\cite{stillinger1988supercooled}. This argument, however, is not conclusive~\cite{eastwood2002droplets,biroli2000inherent} as it only holds when counting potential-energy minima which are much more numerous than metastable free-energy minima in the mean-field limit and are not necessarily relevant for describing the glass transition. We will come back to numerical implementations of this so-defined configurational entropy later on.

A substitute for metastable free-energy states based on collections of liquid configurations could be defined in finite dimensions with the help of the dynamics by setting a threshold for the lifetime of these collections. After all, if the lifetime threshold is large compared to the local equilibration time, the system can equilibrate within the metastable state and metastability is thus observable in practice, as it is the case for the \emph{supercooled} liquid phase with respect to the crystalline phase. The case of glassy metastable states is, however, more subtle because in the observable temperature range all timescales (local equilibration time, lifetime, etc.) are expected to  be comparable. Moreover, while simple symmetry operations can be used to immediately tell apart, {\it e.g.}, supercooled liquid from crystalline phase, there are no such symmetry operations allowing one to distinguish one metastable state from another,  and the very concept of metastable state lifetime is less sharply defined. Configurations can still in principle be grouped into larger entities, on the basis of their relative distance or the height of the energy barriers separating them. Such entities are sometimes referred to as metabasins~\cite{stillinger1995topographic}, but in spite of the pioneering work of A. Heuer and coworkers~\cite{heuer2008exploring}, the procedure is hard to implement numerically and has not been used to discuss the Kauzmann transition~\cite{sastry2004numerical}. 

Another possibility is to restrict the spatial fluctuations so that a nonconvex free-energy landscape with multiple minima can still be defined. This is what happens when considering small system sizes: Long wavelength fluctuations as well as rare localized fluctuations are suppressed by the limited sample size, and some form of metastability can then be studied. A specific protocol of this kind has actually been proposed and implemented to measure a fundamental lengthscale associated with the putative thermodynamic glass transition~\cite{bouchaud2004adam,biroli2008thermodynamic}. By freezing the liquid configuration 
outside a spherical cavity of radius $R$ and letting the system equilibrate inside the cavity in the presence of the frozen boundary condition, one can extract a crossover length corresponding to the cavity size beyond which the boundary condition no longer fixes the state of the liquid inside. In a schematic adaptation of the mean-field theory to finite dimensions, this length results from 
the competition between the interfacial cost due to the surface tension between distinct metastable states and the configurational entropy gain resulting from the multiplicity of available states. It diverges at the Kauzmann transition when the configurational entropy vanishes. Importantly, such a lengthscale,  $\ell_{\rm{PTS}}$, 
which is associated with a point-to-set spatial correlation~\cite{montanari2006rigorous}, can be probed in finite-dimensional liquids, even if the concept of metastable state is not properly defined in the unconstrained (bulk) liquid: see also below. 

An alternative approach is to introduce an appropriate order parameter and the associated free energy that can play the role of the Landau effective potential in the mean-field description while being generalizable to finite dimensions. A convenient choice is the so-called Franz-Parisi 
potential~\cite{franz1995recipes,franz1997phase}. The first step is to introduce an overlap order parameter $Q$ which measures the similarity between two liquid configurations, or, more properly, between the associated coarse-grained density profiles. Considering two configurations of the liquid,   
${\bf r}^N_\alpha \equiv \{ {\bf r}_i^{(\alpha)} \}_{i=1, \dots, N}$ defined from the particle positions ${\bf r}_i^{(\alpha)}$ with $\alpha = 1$, 2, the overlap function between 
the configurations can be defined as
\begin{equation}
\begin{aligned}
Q_a[{\bf r}^N_1, {\bf r}^N_2]=\frac{1}{N}\sum_{i, j=1}^N w (\vert {\bf r}_i^{(1)}-{\bf r}_j^{(2)}\vert/a ),
\end{aligned}
\label{eq_overlap}
\end{equation}
with $w(x)\approx1$ if $x< 1$ and $\approx 0$ otherwise, $a$ being a tolerance associated with the typical amplitude of thermal vibrations in the liquid. As such the overlap function $Q_a$ is small if the configurations are uncorrelated and large if they are strongly correlated, and it can therefore be used to distinguish liquid (low overlap) and glass (high overlap) phases. 

The Franz-Parisi potential represents the average cost to maintain equilibrium liquid configurations ${\bf r}^N$ at a global overlap value $Q$ with a reference liquid configuration ${\bf r}^N_0$. When all configurations are sampled from the equilibrium measure at the same temperature $T$, its expression reads
\begin{equation}
\begin{aligned}
V(Q)= - \frac{T}{N} \int d{ \bf r}^N_0 \frac{e^{-\beta \mathcal H[{\bf r}^N_0]}}{Z}\ln \int d {\bf r}^N \frac{e^{-\beta \mathcal H[{\bf r}^N]}}{Z}
\delta(Q- Q_a[{\bf r}^N, {\bf r}^N_0]),
\end{aligned}
\label{eq_FPpotential}
\end{equation}
where $\beta=1/(k_B T)$, $Z \equiv \int d {\bf r}^N \exp(- \beta \mathcal H[{\bf r}^N])$ is a partition function, $\mathcal H[{\bf r}^N]$ is the liquid 
Hamiltonian, and $\delta(x)$ is the Dirac delta function.

The mean-field Franz-Parisi potential $V(Q)$ loses convexity at some temperature $T_{\rm c}$ and, below the dynamical transition temperature $T_d<T_{\rm c}$, a second minimum that is metastable with respect to the stable liquid minimum appears at a high value of the overlap and corresponds to the glass phase. 
As $T$ further decreases the second minimum becomes deeper and at the Kauzmann temperature $T_K$ the two minima have the same free-energy value: This corresponds to a discontinuous transition between the liquid and the ideal glass. Between $T_d$ and $T_K$ the difference in free-energy between the two minima is exactly the configurational entropy (times the temperature). Note also that below $T_{\rm c}$, tilting the potential by applying a source term linearly coupled to the overlap, $-\epsilon Q$, induces for some $\epsilon^*(T)>0$ a first-order transition between a low-overlap phase and a high-overlap one~\cite{franz1995recipes,franz1997phase}.

\begin{figure}
\centering
\includegraphics[width=0.6\linewidth]{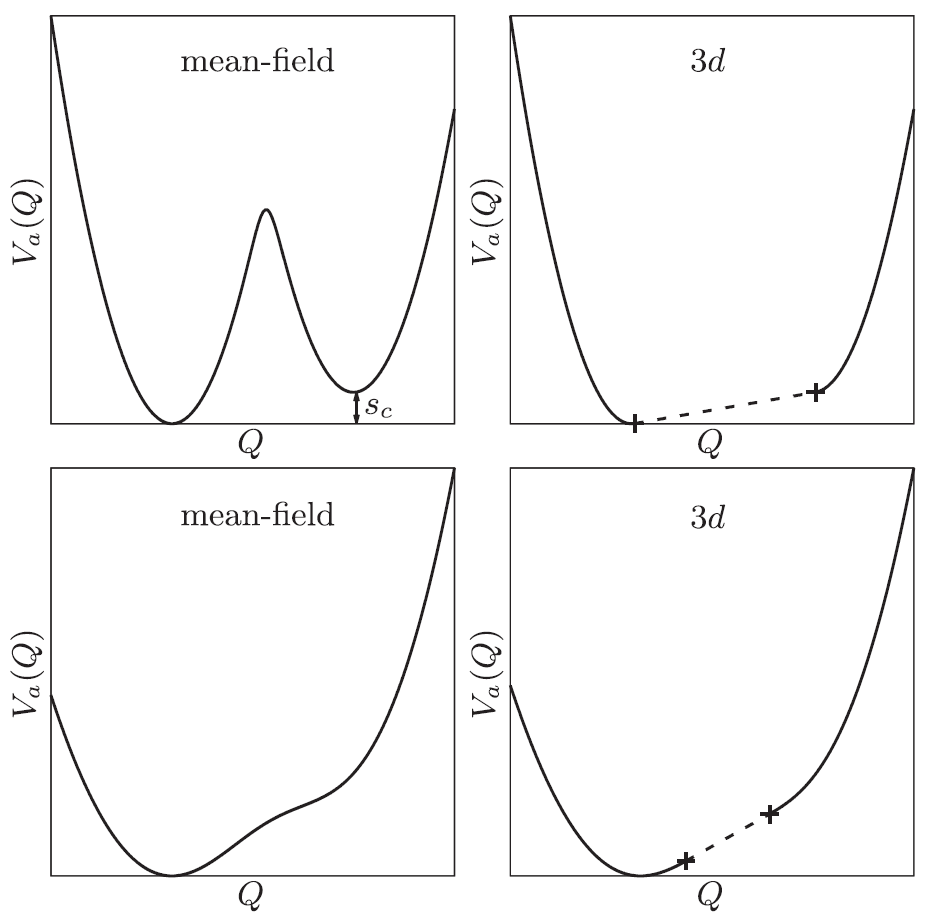}
\caption{Sketch of the Franz-Parisi potential $V(Q)$ in the mean-field limit (left panels) and for a 
3-dimensional glass-former in the thermodynamic limit (right panels). In the top panels, the temperature is slightly above the Kauzmann temperature $T_K$. In the bottom panels the temperature is slightly below $T_{\rm c}$ at which a singular point with $V''(Q) = V'''(Q) = 0$ exists. (Above $T_{\rm c}$ the potential is convex with $V''(Q) > 0$ everywhere.) The figure  is taken from Ref.~\cite{guiselin2020overlap}.}
\label{fig_FP-MF/finite-d}
\end{figure}

The Franz-Parisi potential $V(Q)$ is well-defined in finite dimensions too. It should then be convex in the thermodynamic limit but can nonetheless be 
singular by displaying at low enough temperature a linear segment over a range of $Q$, as illustrated in Fig.~\ref{fig_FP-MF/finite-d}. If so, a discontinuous transition (associated with a horizontal segment) can still take place at some nonzero $T_K$. The difference between the values of $V(Q)$ at the high-overlap end of the segment and at the stable liquid minimum corresponds also in finite dimensions to the difference in free-energy of the two phases and operationally defines a configurational entropy, 
as does in an essentially equivalent way the slope $\epsilon^*(T)$ of the segment. With such definitions, $T_K$ corresponds to a vanishing of the configurational entropy, although the precise link between this configurational entropy and "metastable states" is blurred in finite dimensions.
In any case, one expects that the hypothetical thermodynamic transition in finite dimensions would still be associated with a divergence of the point-to-set correlation length $\ell_{\rm{PTS}}$, as the cavities of the point to set construction would then end up having a vanishing free energy gain in changing state at $T_K$. Although nothing guarantees that the 1-RSB nature of the ideal glass phase persists (see Chapter by P. G. Wolynes), the divergence of the point-to-set correlation length conveys the idea that the glass phase would have an infinite coherence length. In other words, some form of infinite-range "amorphous order" is present~\cite{kurchan2010order} despite the fact that "by naked eye" nothing in the structure seems to distinguish glass from liquid.

One issue that we will now discuss is whether one indeed finds in $3d$ and $2d$ glass-forming liquids, either in finite-size samples or in the thermodynamic limit, an effective potential that looks like the sketches displayed in Fig.~\ref{fig_FP-MF/finite-d} and whether the data is compatible with the presence of an equilibrium glass transition at a nonzero $T_K$.

\section{Does a thermodynamic glass transition exist in $2$ and $3$-dimensional liquids?}
\label{sec_existence}

\subsection{The issue of timescale and system size}

As already stressed, the Kauzmann thermodynamic glass transition is inaccessible by simply cooling a liquid. This is the obvious consequence of the slowing down of relaxation that, in practice, prevents equilibration below some temperature $T_g>T_K$. This is true in experiments~\cite{ediger1996supercooled}, and even alternative ways of 
generating ultrastable glasses by physical vapor deposition~\cite{swallen2007organic,ediger2017perspective,rodriguez2022ultrastable}, fall short of reaching the very near vicinity of the 
extrapolated $T_K$. The situation is even less favorable for computer simulations that span a less extended timescale domain in equilibrium than experimental techniques. However, computer simulations allow one to access microscopic properties in great detail and, more importantly, to implement protocols and compute quantities, such as the point-to-set correlation length, the statistics of the overlap between configurations, the Franz-Parisi potential, etc., that are hard or impossible to probe experimentally.

Furthermore, advanced simulation techniques such as parallel tempering~\cite{marinari1992simulated,hukushima1996exchange} and swap Monte Carlo~\cite{grigera2001fast} which use 
nonphysical particle moves have tremendously increased the range of temperature over which equilibrium liquid configurations can be numerically 
prepared~\cite{berthier2014novel,ninarello2017models,berthier2022modern}. In specifically tailored polydisperse glass-forming models, the 
swap algorithm allows one to sample equilibrium configurations at temperatures below the estimated calorimetric glass transition temperature 
$T_g$~\cite{ninarello2017models,berthier2017configurational,parmar2020ultrastable,berthier2019efficient}. Yet, as in the case of the experimentally generated ultrastable glasses, the close vicinity of the putative $T_K$ is still not attainable (and studying the equilibrium dynamics via physical particle moves remains, of course, limited). This implies that no {\it direct} evidence of a thermodynamic glass transition at $T_K$ can be obtained from such numerical studies at present.

\begin{figure}
\centering
\includegraphics[width=0.6\linewidth]{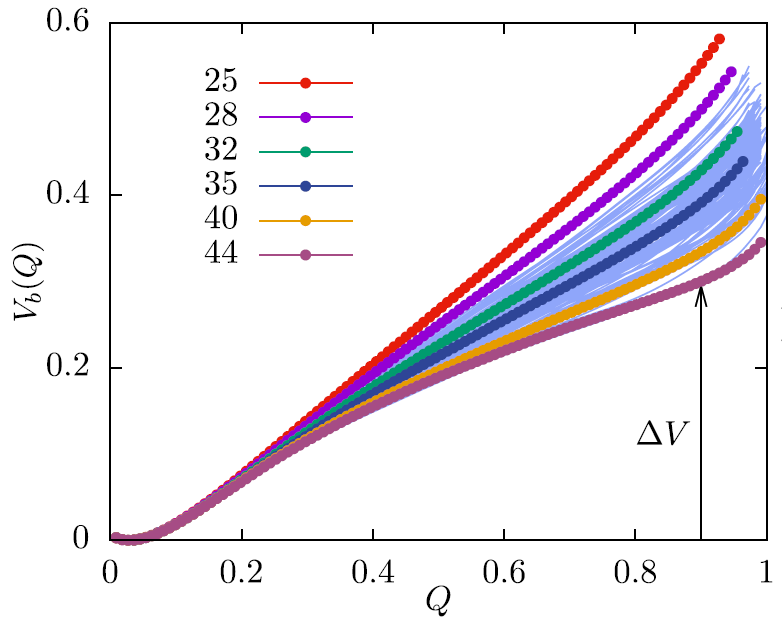}
\caption{Evolution with the pressure of the Franz-Parisi potential $V(Q)$ for a $3d$ polydisperse hard-sphere glass-former with $N=111$ atoms. For pressure $P\gtrsim 28$, the potential is clearly nonconvex, and $\Delta V$ can be taken as an ersatz for a finite-size configurational entropy. 
The figure is taken from Ref.~\cite{berthier2021self}.}
\label{fig_FP-finitesize}
\end{figure}

Another limitation of computer simulations of glass-forming liquids is the accessible range of system sizes. When dealing with the complex computation of quantities associated with the thermodynamic fluctuations of the overlap between configurations at the lowest temperatures at which equilibration is achievable, no more than a few thousands of particles in $3$ dimensions can be simulated with present-day computer capabilities~\cite{ozawa2015equilibrium,guiselin2022statistical}. The thermodynamic limit can then only be inferred through finite-size scaling analyses.
On the other hand, and as already mentioned, the upside of studying systems of limited size is that metastability can be observed and that an estimate of a (finite-size) configurational entropy can be extracted from the measured Franz-Parisi potential. This is illustrated in Fig.~\ref{fig_FP-finitesize} for a system of rather small size~\cite{berthier2021self}. From simple renormalization-group arguments~\cite{cammarota2011renormalization}, one expects that such a finite-size configurational entropy is meaningful in terms of metastable states up to sizes of the order of the point-to-set correlation length. 
For larger system sizes, a configurational entropy can still be operationally defined from the Franz-Parisi potential (even in the thermodynamic limit where convexity is restored, see above and Fig.~\ref{fig_FP-MF/finite-d}), but we stress again that it lacks a direct interpretation in terms of metastable states.

\subsection{Configurational entropy and point-to-set correlation length}

Keeping in mind that the putative Kauzmann transition cannot be directly observed nor very closely approached when decreasing the 
temperature, it is nonetheless worth checking if the behavior of glass-forming liquids is at least compatible with the scenario of glass formation controlled by an underlying thermodynamic glass transition. The two most obvious observables that are relevant for this purpose are the configurational entropy, or rather its different substitutes and estimates, and the point-to-set correlation length extracted from the cavity construction previously discussed. 

The configurational entropy has been studied in $2$- and $3$-dimensional glass-forming liquid models through both the Franz-Parisi potential and the number of inherent structures in the potential-energy landscape~\cite{berthier2014novel,berthier2017configurational,guiselin2022statistical,sengupta2012adam}. For the latter, it is, of course, impossible to proceed by brute-force 
enumeration of the minima except for very small systems~\cite{gao2006frequency,wales2003energy}. The "configurational entropy" is instead approximated by the difference between the full thermodynamic entropy and the vibrational entropy, whose calculation is itself nontrivial due to anharmonicity effects and to the mixing entropy present in polydisperse 
systems~\cite{sciortino2005potential,ozawa2017does,ozawa2018configurational}.  

\begin{figure}
\centering
\includegraphics[width=0.5\linewidth]{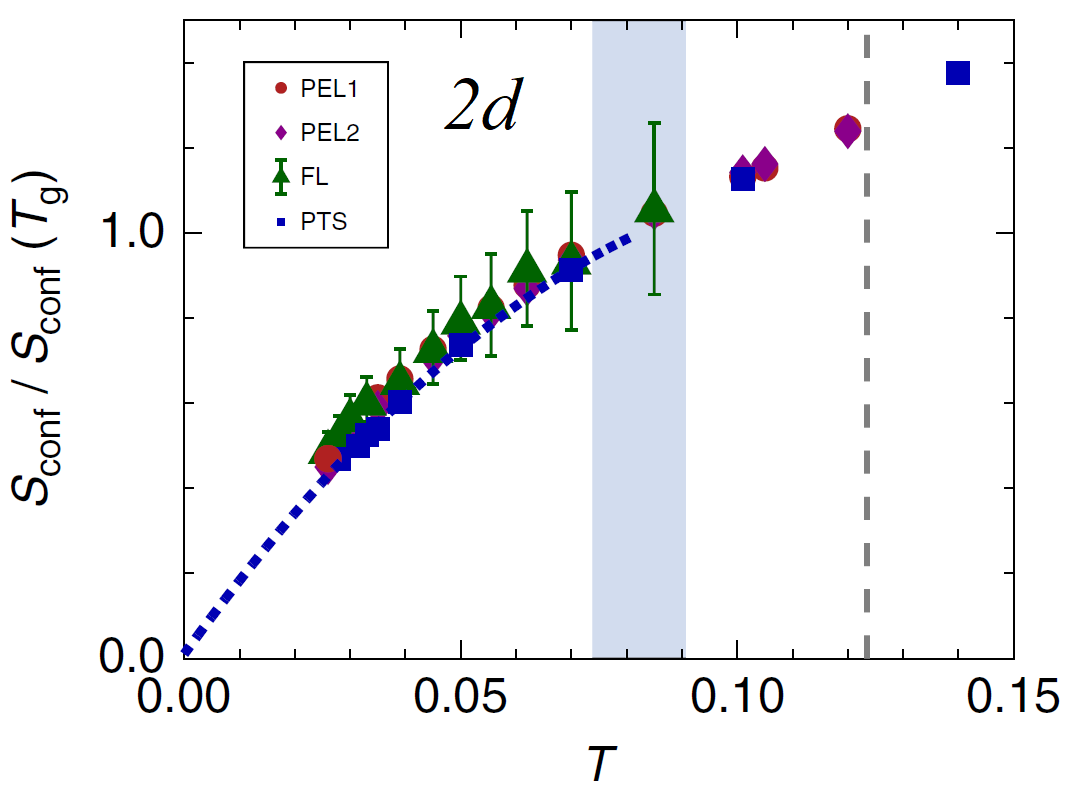}
\includegraphics[width=0.45\linewidth]{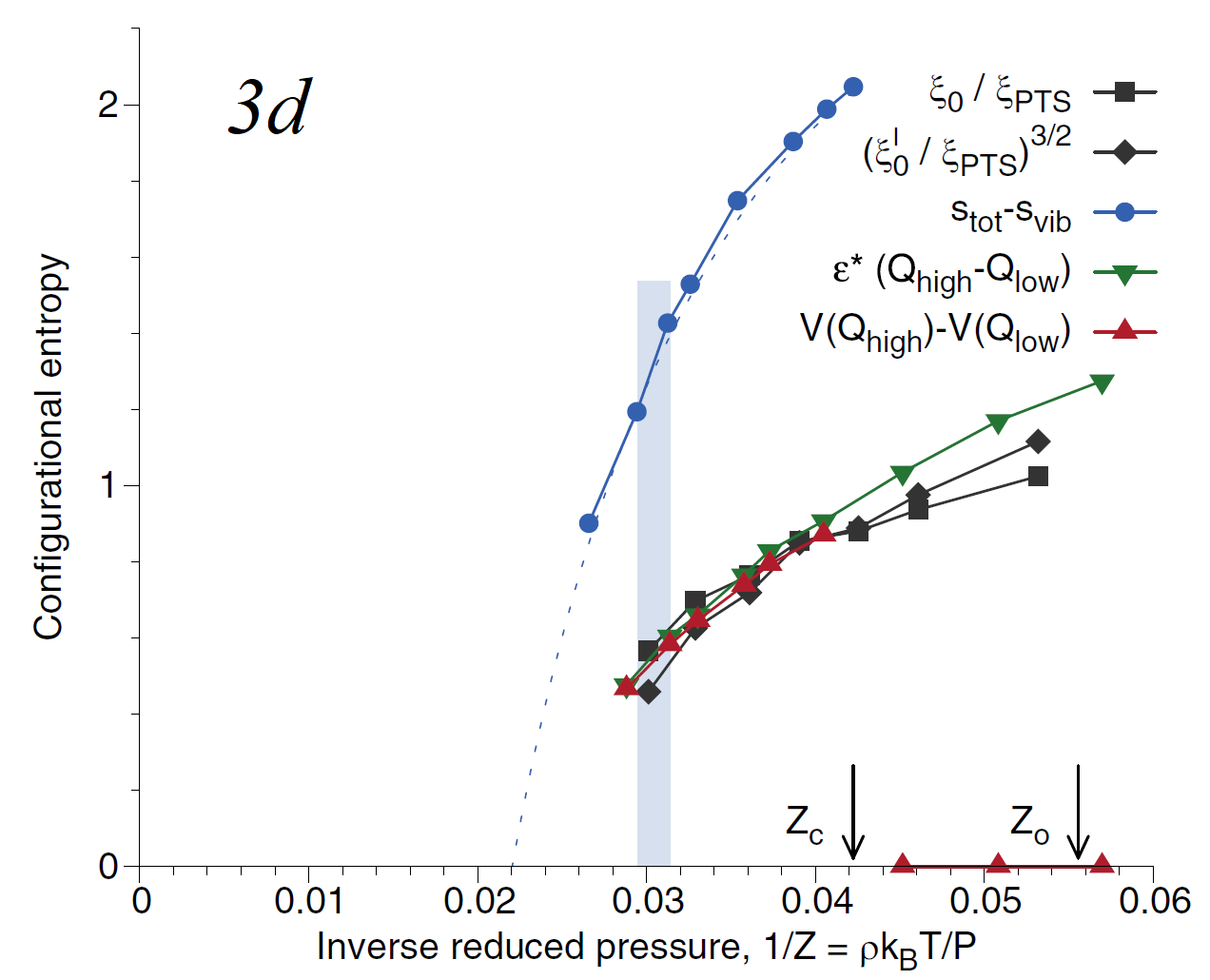}
\caption{Left: Rescaled configurational entropy  obtained through the potential-energy landscape and  the point-to-set correlation length (see main text), for a $2d$ polydisperse soft particles as a function of temperature. The figure is taken from Ref.~\cite{berthier2019zero}. Right: Configurational entropy obtained through the potential-energy landscape, the point-to-set correlation length, and the Franz-Parisi potential (see main text) for a $3d$ polydisperse mixture of hard spheres as a function of inverse reduced pressure. The figure is taken from Ref.~\cite{berthier2017configurational}.
} 
\label{fig_epsT2d3d}
\end{figure}

As anticipated, the configurational entropy associated with the potential energy landscape is significantly larger than that estimated from the Franz-Parisi potential~\cite{berthier2014novel,berthier2017configurational,berthier2019zero}. In all studied cases, the configurational entropy, whatever its definition, decreases with decreasing temperature~\cite{sciortino1999inherent,sastry2000evaluation,berthier2019configurational}, 
much like the experimentally determined one~\cite{tatsumi2012thermodynamic} and Kauzmann's original plot~\cite{kauzmann1948nature}. Different behaviors are found for $2$- and $3$-dimensional glass-forming liquids, as shown in Fig.~\ref{fig_epsT2d3d}.
For $3$-dimensional liquids, extrapolation of the temperature dependence is compatible with a nonzero Kauzmann temperature~\cite{berthier2017configurational,berthier2019configurational} 
whereas for $2$-dimensional glass-forming liquids, $T_K=0$ appears to best fit the data~\cite{berthier2019zero}.

The point-to-set correlation length $\ell_{\rm{PTS}}$ has also been measured by several groups and found to increase with decreasing temperature, by a 
modest factor of $2-3$ in $3d$~\cite{biroli2008thermodynamic,hocky2012growing,berthier2016efficient,berthier2017configurational} and of $6-7$ in 
$2d$~\cite{berthier2019zero} over the accessible temperature range. Again, the numerical data do not approach the anticipated singularity at $T_K$ where $\ell_{\rm{PTS}}$ should diverge and therefore do not prove the existence of an underlying thermodynamic glass transition, but they do show significant growth, much larger than that observed for any simple (point-to-point) structural correlation length. Furthermore, $\ell_{\rm{PTS}}(T)$ behaves as the inverse of the configurational entropy measured from the Franz-Parisi potential~\cite{berthier2017configurational,berthier2019zero} as predicted by the mean-field 
theory (assuming that the surface tension between glassy states essentially does not vary with temperature in the relevant range).

\subsection{A necessary condition for the existence of a Kauzmann transition}

\begin{figure}
\centering
\includegraphics[width=0.95\linewidth]{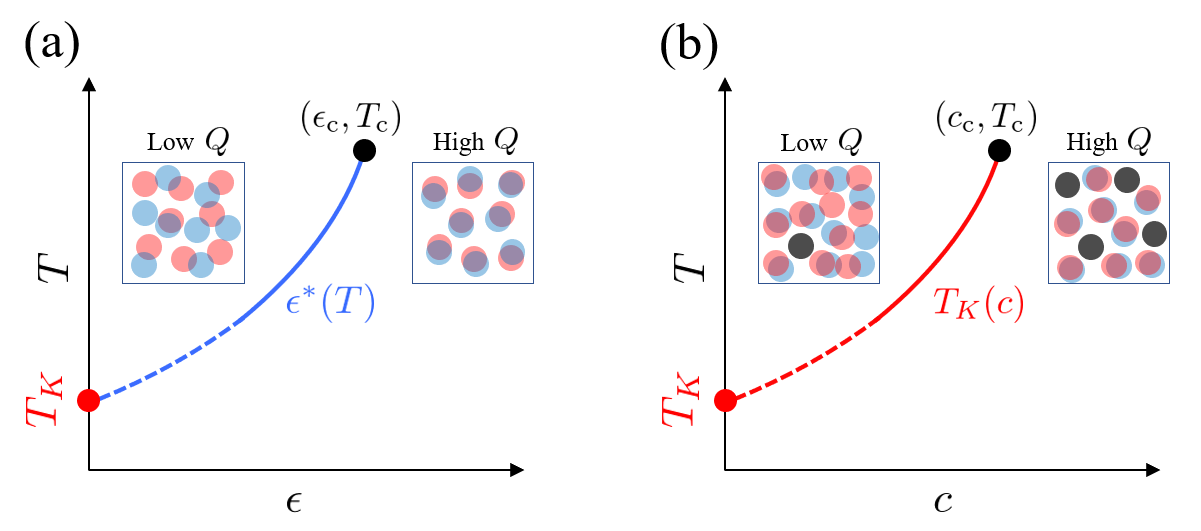}
\caption{Schematic plot of the extended phase diagrams of glass-forming liquids obtained by applying a uniform source $\epsilon$ biasing the value of the overlap $Q$ (a) or by randomly pinning a fraction $c$ of particles (b), expected for 3-dimensional systems. The mean-field theory predicts a line of conventional first-order transition (with latent heat associated with a discontinuous jump of entropy) emerging from the Kauzmann transition in $\epsilon=0$ and ending at a critical point at
$(\epsilon_{\rm c}, T_{\rm c})$ in the former case (a) and a line of Kauzmann transition (at which the configurational entropy vanishes and the point-to-set correlation length diverges) also ending in a critical point at $(c_{\rm c},T_{\rm c})$ in the latter case (b).}
\label{fig_phase-diagrams}
\end{figure}

As already pointed out, despite the overall consistency between numerical results and mean-field predictions, it is virtually impossible to provide a direct experimental or numerical evidence for the  existence of a Kauzmann transition. One may wonder if, conversely, it could be possible to prove that it is absent in finite dimensions. We have already discussed in Sec.~\ref{sec_nature} that {\it a priori} no rigorous arguments seem to preclude the presence of a random-first-order ideal glass transition at a nonzero temperature. An indirect way of addressing the issue is to consider the 
Franz-Parisi effective potential characterizing the fluctuations of the overlap with a reference liquid configuration and the extended phase diagram obtained by applying a linear uniform source $\epsilon>0$ to the overlap: see Sec.~\ref{sec_nature}. In the $\epsilon$ - $T$ phase diagram, the singular behavior of the effective potential, if present, leads to a line of first-order transition ending in a critical point at $T_{\rm c}$ (and a nonzero $\epsilon_{\rm c}$). This is sketched in
Fig.~\ref{fig_phase-diagrams}(a).

\begin{figure}
\centering
\includegraphics[width=0.7\linewidth]{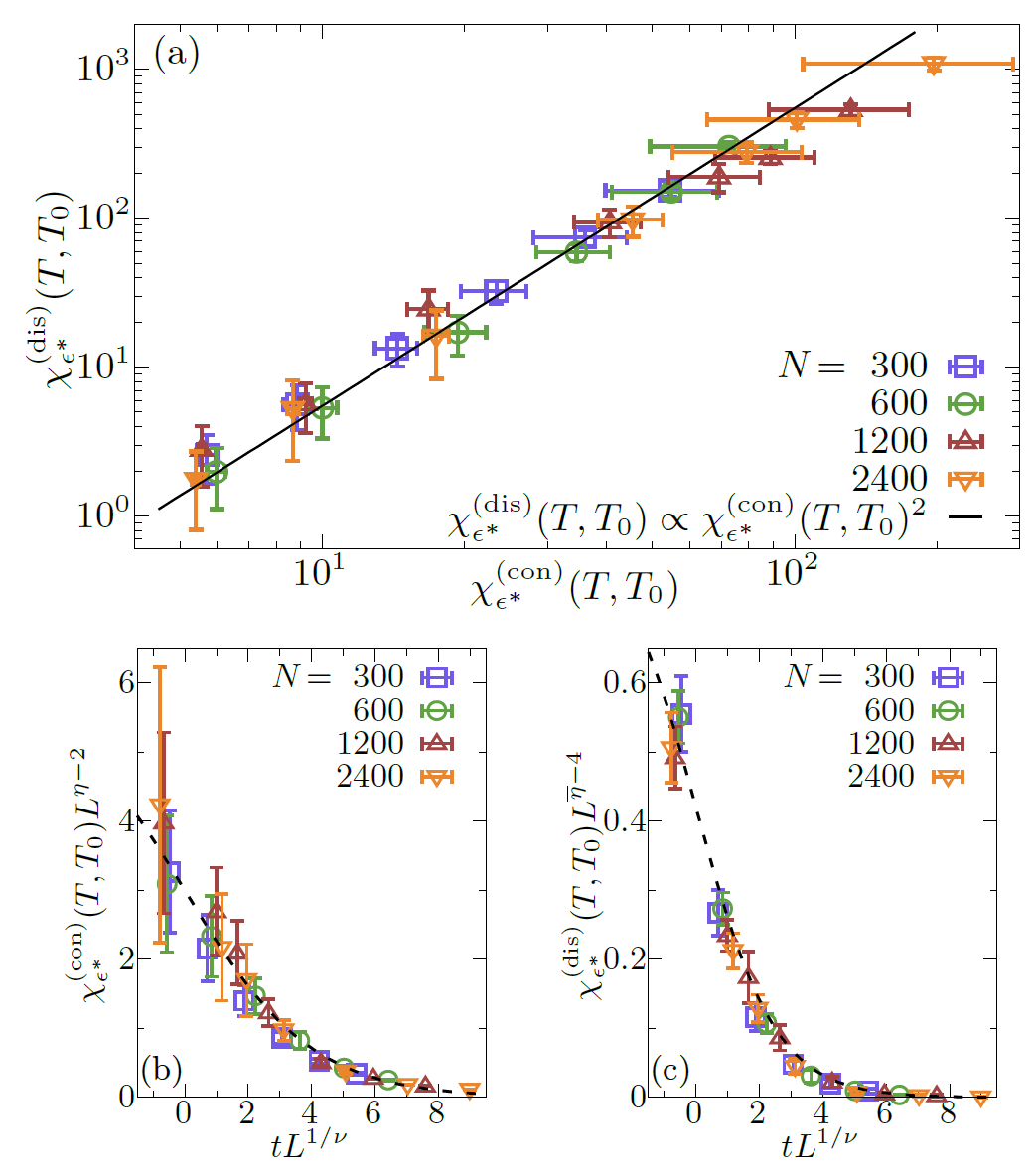}
\caption{Finite-size scaling analysis of the susceptibilities close to the critical point of a
$3$-dimensional glass-forming liquid in the $\epsilon - T$ phase diagram, varying the linear system size $L \propto N^{1/3}$ and the reduced temperature $t=T/T_{\rm c}-1$. The so-called connected and disconnected susceptibilities, $\chi^{\rm (con)}_{\epsilon^*}$ and $\chi^{\rm (dis)}_{\epsilon^*}$, are associated with the
overlap fluctuations, and the scaling collapses are obtained with the exponents, $\eta, \bar \eta, \nu$, characterizing the universality class of the RFIM. The figure is taken from 
Ref.~\cite{guiselin2022statistical}.}
\label{fig_critical-scaling}
\end{figure}

The hypothetical $T_K$ is at the low-temperature limit of the line of first-order transition when $\epsilon=0$. Clearly, if the line does not exists, and in particular, if no critical point exists at some $T_{\rm c}>0$, no Kauzmann transition can be present. Investigating the critical point and the high-temperature part of the first-order line is numerically very demanding but doable~\cite{berthier2013overlap,berthier2014novel,berthier2015evidence}. Computing the Franz-Parisi potential requires sampling a range of large overlap values that correspond to rare 
occurrences, but this can be achieved by using the large-deviation framework and importance sampling techniques~\cite{frenkel2001understanding}. Furthermore, when studying the existence and the nature of a critical point, one can optimize the calculation by playing with the temperature $T_0$ at which the reference configurations are 
sampled~\cite{guiselin2022statistical}. Although the accessible system sizes are limited (see above), a finite-size scaling study has given strong support for the existence of a critical point at a nonzero $T_{\rm c}$ in a $3$-dimensional glass-forming liquid and the absence of such a nonzero $T_{\rm c}$ in 
a $2$-dimensional one~\cite{guiselin2022statistical,guiselin2020random}. As theoretically predicted from field-theoretical 
arguments~\cite{franz2013glassy,franz2013universality,biroli2014random}, the results are compatible with the critical point $(\epsilon_{\rm c}, T_{\rm c})$ being in the universality class of the 
equilibrium random-field Ising model (RFIM) (see Ref.~\cite{nattermann1998theory} for an introduction to the model). In particular, the lower critical dimension of 
the RFIM below which there is no phase transition is rigorously known to be $d=2$. The finite-size scaling analysis for the $3d$ case 
is reproduced in Fig.~\ref{fig_critical-scaling}.

The above study concerning the extended $\epsilon-T$ phase diagram and the associated critical point supports the conclusion that a Kauzmann-like
glass transition is precluded in $2$-dimensional glass-formers while its existence is possible (but not guaranteed) in $3$ dimensions where it may then depend on nonuniversal liquid properties producing the strength of the emergent random field that appears in the effective description: see Ref.~\cite{biroli2014random} and below.
\\

\subsection{A thermodynamic glass transition by random pinning}

The Kauzmann transition is not the only inaccessible transition in statistical physics. The equilibrium paramagnetic-to-ferromagnetic transition of the RFIM, which we have already mentioned, is another example. The slowing down of relaxation when approaching the critical point is anomalously strong and has the very same activated character as the glassy slowdown~\cite{villain1984nonequilibrium,fisher1986scaling}. As a result, it is not possible to reach the near vicinity of the transition point by varying the temperature because the system falls off equilibrium at some point where equilibration is no longer achievable in practice. In some experimental realizations of the RFIM, it is nonetheless possible to approach close enough to observe a critical scaling behavior~\cite{belanger1998experiments}. Furthermore, by making use of an additional control parameter, e.g., the concentration of impurities for the diluted anti-ferromagnet in a uniform magnetic field that is supposed to be in the universality class of the RFIM~\cite{belanger1998experiments}, 
one can access the ordered low-temperature phase. Hence, the transition can in principle be studied by both cooling {\it and} heating the system.

A similar protocol for studying the Kauzmann transition was proposed for glass-forming systems by G. Biroli and one of the authors~\cite{cammarota2012ideal,cammarota2013random}. The idea is to choose at random a fraction $c$ of particles in an equilibrium liquid configuration and to freeze them permanently while studying the further evolution of the remaining particles in the presence of the pinned ones~\cite{kim2003effects}. This corresponds to a different boundary condition than the cavity protocol used to access the point-to-set correlation length. The moving particles now form a continuously connected $3$-dimensional system (in a $3d$ sample, provided that the concentration $c$ of pinned particles is below the percolation threshold), and phase transitions are therefore possible in the thermodynamic limit. The mean-field theory~\cite{cammarota2012ideal} predicts a line of thermodynamic glass transition emerging from the Kauzmann temperature $T_K$ in $c=0$ and ending in a critical point at some $(c_{{\rm c}}>0, T_{{\rm c}}>T_K)$, also argued to be in the RFIM universality class~\cite{franz2013universality,cammarota2013random}. The extended phase diagram in the $c - T$ plane thus appears similar to that previously considered in the $\epsilon - T$ plane, with one crucial difference: The thermodynamic transition all along the line, except at $T_{\rm c}$, now corresponds to a random first-order or Kauzmann-like glass transition, at which the configurational entropy vanishes (yet without replica symmetry breaking). The two extended phase diagrams are sketched in Fig.~\ref{fig_phase-diagrams}.

One advantage of the random pinning protocol or related ones is that it can be realized in real experiments~\cite{gokhale2014growing,kikumoto2020towards,das2021soft}. A second advantage is that the ideal glass phase is known: When working at constant temperature $T$ by varying the concentration $c$, the ideal glass phase in the presence of pinned particles corresponds to the initial equilibrium configuration obtained before any particle pinning~\cite{scheidler2004relaxation}. Furthermore, it is in principle  accessible by following a path in the $c - T$ plane that does not encounter any phase transition. 

Notwithstanding these favorable conditions, assessing the presence of a Kauzmann transition with the pinning construction has proven hard numerically due to issues of equilibration time (even in the ideal glass) and of a proper definition of a configurational entropy~\cite{ozawa2018ideal}. Evidence for a glass transition at some $T_K(c)$ has been obtained in rather small systems and without a  systematic finite-size scaling analysis~\cite{kob2013probing,ozawa2015equilibrium}. There have been studies of the properties of the ideal glass (or a stable glass) in the presence of pinning, giving a first description of its equilibrium fluctuations~\cite{ozawa2018ideal} and its melting into the liquid through nucleation of the latter phase when temperature is raised~\cite{hocky2014equilibrium}. Still, the formation of a glass by nucleation, also predicted by the mean-field theory, has not  been observed due to the too large relaxation times involved (while remaining clear of the percolation threshold for the pinned system). Finally, although its properties should be  qualitatively similar to those of the $(\epsilon_{\rm c},T_{\rm c})$ critical point recently investigated (see above), a systematic study of the critical endpoint $(c_{\rm c},T_{\rm c})$ has not been attempted yet. It is anticipated to be very demanding numerically due to its location at the end of a Kauzmann glass transition line, at which the relaxation time is expected to diverge extremely strongly. On the other hand, establishing the presence of this critical point would more directly confirm the existence of a Kauzmann glass transition.

\section{Conclusion and perspective}

To conclude this chapter, we would like to come back to the issue of why one should bother about the existence of an inaccessible phase transition, which, after all, seems a valid question. One lesson learned from studies on the RFIM in which the transition is also not directly reachable by varying the temperature because of a strong activated-like critical slowing down is that establishing the existence and the properties of the transition via mathematically rigorous methods, the functional renormalization group,  or specific (unphysical) algorithms at zero temperature (see Ref.~\cite{tarjus2020random} for a review), allows one to rationalize the whole phenomenology observed experimentally and numerically, to cook up adapted observables, and when possible to perform scaling collapses of data. Reaching the same state-of-the-art would clearly be a major step forward for the study of glass-forming liquids.

We have seen that as far as the statics is concerned, the observables devised within the mean-field setting to reflect the properties of the underlying 
free-energy landscape can be extended to finite dimensions and that they display an overall behavior that is compatible with the existence of a thermodynamic glass transition in $3d$~\cite{berthier2017configurational,guiselin2020random,ozawa2015equilibrium}, but not in $2d$~\cite{berthier2019zero,guiselin2022statistical}. This, of course, does not prove the existence of a Kauzmann, or random first-order, transition in $3$-dimensional liquids, but it validates the mean-field scenario as a reasonable starting point for theoretical developments. The concepts of metastable free-energy states and configurational entropy should be bypassed or strongly modified when dealing with finite dimensions. Yet, as discussed above, no fundamental arguments seem to forbid the existence of a thermodynamic glass transition in $3$-dimensions (whereas there are such arguments against the existence of a nonzero $T_K$ in $2d$). The existence or not of a transition is therefore a nonuniversal feature that may depend on the specific properties of the $3d$ glass-former. (Indeed, one can construct glass-forming models without a Kauzmann transition~\cite{moreno2006non,smallenburg2013liquids,xu2016generalized,parisi2020theory}.) This also opens the possibility of finding glass-formers with a narrowly avoided thermodynamic glass transition, which for all practical purposes, would play 
the same role as a true transition. A path to address this issue is via the development of an effective theory taking the overlap between configurations as the fundamental field or variable. Work in this direction suggests that such an effective theory involves {\it quenched disorder} in the form of a random field and a random 
coupling~\cite{stevenson2008constructing,biroli2018random1,biroli2018random2} and that the existence or not of a transition (or how narrowly it is avoided) primarily depends on the relative strength of the random field. Extending the numerical investigation of the global Franz-Parisi potential to that of a local version may then allow one to determine the parameters entering in the effective theory~\cite{guiselin2022static}. A similar endeavor could be pursued for the random pinning construction in which quenched disorder plays an important role as well.

More difficult at present is establishing that the dynamical slowdown of relaxation leading to glass transformation is indeed controlled by a thermodynamic glass transition (even a narrowly avoided one). The glassy slowing down prevents an exploration of the asymptotic regime in which scaling about $T_K$ is dominant so that exponents cannot be reliably extracted. Note that this is also partly true in the case of the RFIM, for which the range of dynamical data is not sufficient to accurately determine the exponent characterizing the growth of the activation free-energy barriers when approaching the critical point. However, what is more crucial for glass-forming liquids is that room is left for important if not predominant contributions coming from other dynamical processes unrelated to $T_K$, such as dynamical facilitation~\cite{chandler2010dynamics}, soft modes~\cite{widmer2008irreversible}, elasticity~\cite{dyre2006colloquium}, or the role of liquid-specific locally preferred atomic arrangements~\cite{royall2015role,tanaka2010critical}. Establishing the causal relationship between the 
putative thermodynamic glass transition at $T_K$ and the glassy dynamics beyond the existing empirical but global correlations~\cite{lubchenko2007theory} is  therefore the most wanted next stage~\cite{biroli2022rfot}.

\bibliographystyle{ws-rv-van}


\printindex                         

\end{document}